# Giant magnetoresistance in single layer graphene flakes with a gate voltage tunable weak antilocalization

Kalon Gopinadhan,[1,2] Young Jun Shin,[1] Indra Yudhistira,[2,3] Jing Niu,[1] and Hyunsoo Yang[1,2,*]

[1]Department of Electrical and Computer Engineering and NUSNNI, National University of Singapore, 117576 Singapore

[2]Graphene Research Centre, National University of Singapore, 117546 Singapore

[3]Department of Physics, National University of Singapore, 117542 Singapore

**ABSTRACT**

A clear gate voltage tunable weak antilocalization and a giant magnetoresistance of ~ 400 % are observed at 1.9 K in single layer graphene with an out-of-plane field. A large magnetoresistance value of 275% is obtained even at room temperature implying potential applications of graphene in magnetic sensors. Both the weak antilocalization and giant magnetoresistance persists far away from the charge neutrality point in contrast to previous reports, and both effects are originated from charged impurities. Interestingly, the signatures of Shubnikov–de Haas oscillations and the quantum Hall effect are also observed for the same sample.

Graphene, the thinnest electronic material in two dimensions, is a monolayer of carbon atoms arranged at the six corners of a hexagonal honeycomb lattice. When compared to conventional two dimensional materials, monolayer graphene has a linear band structure at low energies which implies a zero electron (hole) effective mass.[1-3] In addition, a very weak momentum scattering from phonons leads to a high mobility for the carriers. For example, a very large mobility approaching 200,000 cm$^2$V$^{-1}$s$^{-1}$ at 4 K has been reported in suspended single layer graphene,[4] however, the mobility is reduced on a supported substrate due to scattering from phonons and charged impurities of the substrate surface.[5] The electric field tunability of its ambipolar characteristic makes them very unique from other two dimensional materials, which is very useful in electronic applications such as frequency multipliers.[6] Even though graphene possesses many unique properties in comparison to Si, the absence of an electronic band gap makes it quite difficult for transistor (digital) applications, but it can still be used in analogue, high frequency applications.

Electron transport studies in single layer graphene in the past, both theoretically and experimentally, reveal many interesting properties such as gate tunable carrier density,[7] minimum conductivity at or near the Dirac point,[8] ballistic conductivity,[9] and relativistic quantum Hall effect with Hall plateaus[2,10]. In addition, charge carriers in graphene possess chirality, and the Berry phase of the carriers in single layer graphene is π. This additional phase results in a weak antilocalization on traveling a closed trajectory rather than a weak localization normally seen in conventional two dimensional systems. Contrary to this, weak localization is routinely observed in graphene systems and it is identified as a result of inter-valley and intra-valley scattering caused by atomically sharp point defects.[11] Weak localization along with an antilocalization is also reported in epitaxial graphene on SiC as an evidence of chiral Dirac fermions.[12] A recent study on mechanically exfoliated graphene reports a transition from weak localization to weak antilocalization upon lowering the carrier density and increasing the temperature.[13] However, a clear weak antilocalization at low temperatures has not been reported in single layer graphene and its tunability with carrier density is not achieved.

Another important technological attribute is the magnetoresistance (MR), the change in the electrical resistance upon application of magnetic fields. For practical applications, a large value of MR at room temperature is necessary and the intrinsic properties of graphene need to be utilized.[14] A large negative magnetoresistance is predicted[15] and observed[16] in graphene nanoribbons. A positive MR up to 120% is observed in multilayer epitaxial graphene[17] and a MR up to 100% is obtained from sandwiched CVD graphene samples.[18] A large MR near the charge neutrality point (CNP) is reported[19] in single layer graphene, however, there has been little report of a large intrinsic MR away from the CNP in single layer graphene at room temperature.

In this paper, we report the observation of a large oscillatory MR up a value of 400 % in single layer graphene flakes at 1.9 K as well as a weak antilocalization quantum correction to the transport as expected for Dirac fermions. A large MR value of 275% persists up to room temperature and in wide range of gate voltage, even away from the CNP. Interestingly, our sample also shows an onset of Shubnikov–de Haas (SdH) oscillations and integer quantum Hall effect (IQHE) at 9 T and 1.9 K, even though there is a signature of defects from the Raman spectroscopy data.

The single layer graphene was prepared by micromechanical exfoliation of Kish graphite followed by a transfer to a highly p-doped Si substrate, which was covered by a layer of 300 nm thick $SiO_2$. Mechanically cleaved graphene was identified by an optical microscope and further quantified by Raman spectrophotometer.[20-22] Electrodes were patterned by e-beam lithography using a combination of MMA/PMMA, and subsequently Cr/Au (7 nm/95 nm) was deposited by an e-beam evaporator, where a large distance between the source materials and the sample keeps the damage induced in graphene at minimal. Standard lift-off procedures using warm acetone were followed after the deposition. The transport measurements were carried out in a Hall bar geometry in a physical property measurement system under He atmosphere. Before the measurement, the samples have been annealed for 2 hours at 400 K under high vacuum conditions to remove any adsorbed water vapor. To apply back gate bias, the source terminal was connected to the back gate and the leakage current through the $SiO_2$ layer was monitored.

The optical microscopy image of the fabricated device is shown in Fig. 1(a). In order to determine the quality and thickness of graphene, Raman spectroscopy measurements were carried out. A laser wavelength of 532 nm with a power density ~ 0.5 mW/cm$^2$ was used to avoid any laser induced heating. Raman spectrum of pristine graphene is shown in Fig. 1(b). The prominent modes in the spectrum are the G mode at 1587 cm$^{-1}$, G* mode at 2456 cm$^{-1}$, and the 2D mode at 2680 cm$^{-1}$. A Lorentzian peak fitting is performed on the 2D peak and a good fit is obtained with a single Lorentzian with a full width at half maximum (FWHM) of ~25 cm$^{-1}$ as shown in Fig. 1(c). The above estimated FWHM agrees well with the reported result for a single layer graphene.[23, 24] A weak disorder (D) peak is visible at ~1340 cm$^{-1}$ which implies the presence of short range defects in the sample. Figure 1(d) shows the conductivity (σ) of single layer graphene as a function of back gate voltage ($V_G$) at 1.9 K. The conductivity is calculated using the relation $\sigma = \frac{1}{R}(\frac{L}{W})$ where $R$ is the sheet resistance, $L$ is the distance between the voltage probes, and $W$ is the width of the graphene channel. The conductivity shows a sub-linear relationship with respect to $V_G$ with a smeared tail near the CNP where the conductivity is at minimum, in contrast to a sharp minimum expected for homogeneous single layer graphene. We observe asymmetry in our experimental data between the electron and hole transport. Using our sample dimensions and the formalism of Huard et al.[25], we can calculate that our sample geometry accounts for equal or less than 2 percent difference between the measured and actual electron and hole mobility. Thus we conclude that the role of invasive probes is negligible and the observed asymmetry is intrinsic in origin. The carrier mobility is calculated using the relation $1/\sigma_{e,h} = 1/|n|e\mu_{e,h} + 1/C$ which are 0.96 m²/(V·s) and 0.27 m²/(V·s) for electrons and holes, respectively, where both $\mu_{e,h}$ and C are fitting parameters. The geometric mean of mobility of 0.51 m²/(V·s) is used later to calculate the charged impurity concentration. To quantify the gate dependent conductivity behavior, we have used a self consistent theory[8] to fit our data and the resultant plot is shown in Fig. 1(d). The fitting yields an impurity density $n_{imp,e}$ = 25.0×10¹⁰ cm⁻² for electron, and $n_{imp,h}$ = 148.3×10¹⁰ cm⁻² for hole, which yields a geometrically averaged impurity density of 72.7×10¹⁰ cm⁻² and short range conductivity $\sigma_s$ = 56.1 e²/h. The estimated average charged impurity concentration is similar to that derived from the Raman spectrum (80×10¹⁰ cm⁻²). The short range conductivity is arising from the sharp potential created by short range scatterers such as point defects and dislocations in the lattice.[8] The short range scattering induces the sub-linear behavior usually observed in the conductivity vs. gate voltage curve at high gate voltages. From Matthiessen's resistivity rule, it is clear that the effect of short range conductivity (56 e²/h) is small at the minimum conductivity point (10 e²/h) which suggests that charged impurity scattering by long range scatterers is dominated at the minimum conductivity point. The CNP occurs at ~20 V which indicates the sample is lightly hole-doped. It is known that adsorption of dopants such as $H_2O$ and $O_2$ make graphene hole-doped, even though the sample has been vacuum annealed at 400 K before the measurements.[26]

Figure 2(a) shows the resistivity (ρ) as a function of $V_G$ from 1.9 to 200 K. The CNP shifts to higher gate voltages as the temperature increases from 1.9 K, and the shift of the CNP can be utilized to estimate the carrier concentration in the sample at different temperatures. In order to estimate the carrier concentration, the capacitance of the SiO$_2$ gate dielectric is utilized. By assuming a SiO$_2$ thickness (*d*) of 300 nm and dielectric constant ($\kappa$) of 3.9, the gate capacitance per unit area, $C_g = \varepsilon_0 \kappa / d$ is $1.15 \times 10^{-8}$ F/cm$^2$ which is very close to the measured capacitance (through Hall measurements) $C_{gm} = 1.36 \times 10^{-8}$ F/cm$^2$. In the capacitor model, the carrier concentration is estimated using the relation, $n = -C_g(V_G - V_D)/e$, where $V_D$ is the gate voltage corresponding to the CNP. For a zero gate voltage, the hole concentration shows a decrease with decreasing temperature from 200 K. At 200 K the estimated hole concentration is $1.29 \times 10^{12}$ cm$^{-2}$, whereas it is $1.16 \times 10^{12}$ cm$^{-2}$ at 1.9 K. The decrease in carrier concentration with decreasing temperature may attribute to the thermal excitation of carriers across a zero-bandgap semiconductor, however, the decrease in carrier concentration is much smaller than that of reported values in literature.[7] Alternatively, such a small change in the carrier concentration may arise from tunneling across random p-n junction networks formed by charged impurities. It is estimated that at zero $V_G$, the hole concentration (*n*) decreases by a factor of 0.90, whereas the conductivity (σ) decreases by a factor of 0.80 upon cooling the sample from 200 to 1.9 K. From the Drude's relationship, mobility $\mu = \sigma / (en)$ decreases by a factor of 0.89 upon cooling the sample from 200 to 1.9 K. It is reasonable to assume that a small change in mobility with temperature is due to charged impurities rather than surface phonons of the SiO$_2$ substrate.[27] The above observation is true independent of the applied gate voltages and it is in line with the reported result, in which charged impurity scattering limits the mobility in single layer graphene.[27]

Figure 2(b) shows the resistivity (ρ) as a function of temperature at different $V_G$. The resistivity shows a saturating characteristic at low temperatures and at all applied gate voltages in contrast to the expected semi-metallic character of graphene. Bilayer graphene is reported to show a similar characteristic due to the existence of electron-hole puddles,[20, 28] however single layer graphene is predicted to be less susceptible to disorder.[29] It is argued that potential fluctuation strength is much weaker in single layer graphene due to a linear density of states. Our result suggests that single layer graphene is equally susceptible to disorder and the resistivity versus temperature characteristic strongly suggests the existence of electron-hole puddles in single layer graphene. The temperature dependent resistivity behavior shown in Fig. 2(b) is predicted theoretically for a charge disordered sample.[30] Near the CNP, the resistivity is temperature dependent, while away from the CNP it is weakly temperature dependent, which is the characteristics of charged impurity dominated transport. The finite temperature dependence is due to the thermal smearing of the Fermi surfaces. Close to the CNP the smearing leads to occupation of both electron and hole bands leading to a larger conductance, while far from the CNP only a single band remains occupied.[30] The temperature dependence is another signature of large electron-hole puddles close to CNP which is important later, when we discuss the magnetoresistance behavior in terms of the population of electron and hole puddles. The strongly distorted conductivity versus gate bias also indicates macroscopic inhomogeniety in the sample. Scanning single electron transistor studies on single layer graphene[31] show an evidence of electron-hole puddles due to charged impurities from the $SiO_2$ substrate. Note that Fig. 2(b) shows that electron-hole puddles also exist away from the Dirac point, which is responsible for the observed large MR away from the CNP as we discuss later.

Figure 2(c) shows the resistivity as a function of $V_G$ at different out-of-plane magnetic fields (H) at 1.9 K. It is clear that on increasing the strength of the magnetic field, the resistivity shows an oscillation characteristic of a two dimensional system. We discuss this in terms of SdH oscillations later. The oscillations are not dependent on the polarity of the magnetic field, however a large increase in resistivity is observed upon an external magnetic field. The resultant MR is quantified as $MR = [\rho(H) - \rho(0)]/\rho(0)$ which is shown in Fig. 2(d) at 9 T as a function of $V_G$. A large positive MR of ~400% is observed near the CNP at 1.9 K and its magnitude decreases on either side of the CNP. The MR also exhibits clear SdH oscillations on top of a large positive background. It is understood that a large positive MR background is a result of inhomogeneous Drude conductivity in the sample.[32] The gate tunability of the MR implies that the intrinsic character of graphene is the source of the observed MR phenomenon.

Figure 3(a) shows the MR versus out-of-plane magnetic fields (an intermediate field regime) as a function of $V_G$ at 1.9 K. The MR shows a sharp cusp at very low magnetic fields and a positive slope at higher fields. The sharp cusp at low magnetic fields suggests a weak antilocalization of the carriers. Note that the maximum out-of-plane magnetic field applied to the sample is only 0.8 T in this graph. The positive MR is a result of the Lorentz force induced deflection of the carriers under a magnetic field. In a homogeneous medium with a single carrier type, there is no transverse magnetoresistivity [$\rho_{xy}(H)$] since the Lorentz force cancels the force due to the Hall electric field and the longitudinal magnetoresistivity [$\rho_{xx}(H)$] is proportional to $1+(\mu H)^2$.[19] For graphene, at the CNP, the existence of both electron and hole carriers can give rise to finite magnetoresistance. Theoretically, there have been two proposals for calculating this effect. Hwang et al.[33] proposed a two-fluid model where the resistivity ($\rho_{xx}$) and the Hall resistivity ($\rho_{xy}$) is given by $\rho_{xx}(H) = \rho_{xx}(0)(1+(\mu H)^2)/(1+(\alpha\mu H)^2)$ and $\rho_{xy}(H) = \alpha\mu H \rho_{xx}(H)$

, respectively, where $\alpha = (n-p)/(n+p)$ with concentrations of electrons ($n$) and holes ($p$) (see also Ref. 19). In Fig. 3(a), the MR has been fitted by defining MR = $(\rho_{xx}(H) - \rho_{xx}(0))/\rho_{xx}(0) = (1-\alpha^2)(\mu H)^2/(1+(\alpha\mu H)^2)$ choosing a constant mobility of 1.3 m$^2$/(V·s) for all the applied gate voltages. The resultant $\alpha$ is plotted as a function of $V_G$ along with a theoretical prediction based on two channel model in Fig. 3(b). Ideally, the value of $\alpha$ is expected to be zero at the CNP, as there is equal concentration of electrons and holes in charge neutral graphene. However, interestingly we rather observe finite values of $\alpha$ at various gate bias voltages including the CNP, which suggests that two channel model is inadequate to accurately explain magnetoresistance.

Tiwari et al.[34] proposed an effective medium theory where the electron-hole puddle induced carrier inhomogeniety gives a magnetoresistance behavior at the CNP. This model takes into account the presence of electron-hole puddles and the distortion of current lines under a magnetic field with the assumption that the size of the electron-hole puddle is larger than the carrier mean free path. This model is only accurate very close to CNP. It is clear from Fig. 2(c) that a large magnetic field driven resistivity enhancement is seen near the CNP, which implies that a large population of electron-hole puddles distorts the current lines, thereby enhances the scattering, resulting in a large magnetoresistance. However, we emphasize that neither the two channel model, nor the effective medium theory explain our large magnetoresistance away from the Dirac point. Therefore, a better theoretical understanding is required to explain the observed large MR away from the Dirac point in an inhomogeneous medium.

Now we discuss in detail the very low magnetic field regime. Figure 4(a) shows the MR data at $V_G$ = 30 V where a sharp cusp at very low magnetic fields is a signature of the weak antilocalization of the charge carriers. Figure 4(b) shows the magnetoconductance at low magnetic fields as a function of $V_G$ at 1.9 K, and the negative magnetoconductance is fitted with a two-parameter empirical model on graphene.[35] The conductance in the presence of a magnetic field $B$ is given by $\Delta\sigma(B) = \frac{e^2}{\pi h}F(\frac{B}{B_\varphi}) - 3F(\frac{B}{B_\varphi + 2B_*})$, where $F(z) = \ln z + \psi(1/2 + 1/z)$, $\psi(x)$ is the digamma function, $B_\varphi$ is the dephasing field, and $B_*$ is an elastic scattering field that has contributions of both inter- and intra-valley scattering. The parameters of the fittings are shown in Table 1. The presence of long range scatterers (charged impurities) that do not distinguish between A and B lattice atoms conserves the pseudospin, therefore there is no back scattering of the charged particles which results in weak antilocalization. From the dephasing field $B_\varphi$, we have extracted the phase coherence length $L_\varphi$ using $B_\varphi = \hbar/4eL_\varphi^2$, as shown in Fig. 4(c). The phase coherence length is gate tunable and it increases with increasing the carrier concentration on either side of the CNP. A coherence length of 185 nm at $V_G$ = -15 V is estimated. Near the CNP (20 V), the coherence length is found to increase to a large value which could be a result of an increase in mobility. The observation of weak antilocalization is significant, as it demonstrates the possibility of a spin-orbit like effect in graphene by introducing charged impurities, which is essential to have gate tunable field effect spin transistors. Gate tunable phase coherence length is reported and estimated in single[36] and bilayer graphene[37] from weak localization features, and our estimates from weak antilocalization features match well with these studies. We have also tried to fit the negative magnetoconductance data with the weak (anti)localization theory[11] (see supplementary materials). The extracted phase coherence lengths are similar, however the phenomenological model provides a better fitting trend for our samples than weak (anti)localization theory.

Figure 4(d) shows the temperature dependent MR from another single layer graphene sample (see supplementary materials). The MR value is also large which is ~275% at 300 K. Most importantly, the giant MR persists even up to room temperature suggesting graphene as a potential candidate in magnetic sensors and read heads of hard disk drives. A nearly temperature independent MR may suggest that the mobility (since the change in carrier concentration is very small in this temperature range) is mostly temperature independent which is similar to the result of zero magnetic field data in Fig. 2(a), providing further evidence of the role of charged impurities in determining MR. The SdH oscillations observed at low temperatures eventually disappear as temperature increases. The observation of giant MR at 300 K rules out any quantum mechanical origin such as weak (anti)localization.[17] In addition, weak antilocalization is easily destroyed with a very small magnetic field (~50 mT), whereas the observed MR persists upto very high magnetic fields. The persistence of giant MR up to 300 K suggests that the positive MR background usually observed at low temperatures is not due to weak antilocalization.[12] In the previous study of weak antilocalization in epitaxial graphene by Wu *et al.*,[12] the observation of temperature independent critical field at which a transition from weak localization to weak antilocalization happens also proves the MR is not due to weak antilocalization, because the dephasing field is a strong function of temperature. Rather, it is related to off-axis Drude conductivity terms arising from inhomogeniety.[38] Electron-electron interaction can also lead to a positive or negative MR,[39] which is highly temperature dependent, whereas in our case the MR is relatively independent of temperature and is always positive.

Figure 5(a) shows the resistivity ($\rho_{xx}$) as a function of $V_G$ at 1.9 K with an out-of-plane magnetic field of 9 T. The resistivity shows signature of Shubnikov–de Haas (SdH) oscillations, and the Hall resistivity in Fig. 5(b) shows signature of IQHE, when $V_G$ is varied. The dips in SdH oscillations and the plateaus in QHE suggest the onset of Landau levels (LL) at 9 T. The Hall resistivity $R_{xy}$ in Fig. 5(b) shows a value of 12.5 kΩ near the CNP which correspond to $0^{th}$ LL on both sides of the CNP. However, a clear plateau indexing is rather cumbersome away from CNP as the width of the plateaus are very small as well as due to a large asymmetry in the electron and hole transport. The non-zero values of $ρ_{xx}$ at the Landau level indicate a classical MR background, in addition to the quantum Hall effects and SdH oscillations. We have mentioned before that a large positive MR background is a result of inhomogeneous Drude conductivity in the sample. It is interesting to see that even though the sample has both charged impurities which break electron-hole symmetry and the atomically sharp defects as inferred from the D-peak of the Raman spectrum, we are able to observe the SdH oscillations and IQHE. It was reported that the defects can enhance the amplitude of SdH oscillations.[12]

Figure 5(c) shows the low magnetic field Hall resistivity as a function of magnetic field at different $V_G$. Across the CNP upon changing $V_G$, the polarity of the slopes changes due to a change in the majority carrier type. Figure 5(d) shows the comparison between the estimated carrier concentration from the Hall resistivity data and the capacitive model using a $SiO_2$ gate dielectric. The magnitude of the carrier concentration is slightly different in these two cases, however the behavior remains the same. This difference strongly points to the presence of charged impurities in the sample which is of the order of $\sim 10^{12}$ cm$^{-2}$ and the difference is also gate voltage dependent. Thus the correction to the carrier concentration is larger near the CNP as shown in Fig. 5(d). We have also tried to estimate the concentration of charged impurities from the Raman spectrum using $I_{2D}/I_G$, where $I_{2D}$ and $I_G$ are the integrated intensity of the 2D and G peak, respectively.[40] For $I_{2D}/I_G = 4.8$, the charged impurity concentration is estimated to be $80\times10^{10}$ cm$^{-2}$ at zero gate voltage which is very similar to that estimated from the fit of conductivity versus gate voltage curve. A similar result has been obtained from another sample (see supplementary materials).

It is clear that there are three regions of interest; one belongs to low magnetic field (few mT) where weak antilocalization is seen, second one to moderate magnetic fields (from 50 mT to 3 T) where the classical MR is seen, and third to high magnetic fields (> 3 T) where SdH, Quantum Hall effect, and classical MR are observed. The observation of a giant MR in single layer graphene opens up practical applications in the field of magnetic sensors. The random p-n junction network based magnetic sensors were reported in silicon,[32, 41] however graphene may offer better performance in terms of cost, mechanical flexibility, and operation temperature. Reports of a giant MR in silicon, silver based chalcogenides,[42] and InSb based disks[43] suggest that current distortions either across a p-n junction or modified geometry[32] can enhance the MR by many folds, and the giant MR can be effectively modeled by the creation of random resistor networks.[38] A giant nonlocality has been reported near the CNP[44] through non-local measurements which may have some influence on the MR, however our measurements are of local in nature and hence its influence on the MR is not very straightforward. We also see a large MR away from the Dirac point which suggests that the giant nonlocality may not be relevant in our case.

In summary, a giant MR of ~ 400 % is observed at 1.9 K in single layer graphene with an applied field of 9 T which is gate tunable and persists even up to 300 K, implying potential applications of graphene in magnetic sensors. The giant MR is explained in terms of the inhomogeneous charge distribution due to charged impurities which creates a random resistor network. A clear gate voltage tunable weak antilocalization is also observed at 1.9 K, supporting charged impurity scattering in our samples. Our observation implies the possibility of spin field effect transistors employing charged impurity scattering as the source of spin-orbit interaction. Signatures of SdH oscillations and the QHE are also seen for the same sample with sharp plateaus in the Hall conductivity.

This work was supported by the Singapore National Research Foundation under CRP Award No. NRF-CRP 4-2008-06. IY is supported by the Singapore National Research Foundation under NRF-NRFF2012-01. We thank Shaffique Adam for helpful discussions.

* Electronic address: eleyang@nus.edu.sg

**Figure Captions**

**FIG. 1.** (a) Optical micrograph of the patterned graphene device. (b) Raman spectrum of single layer graphene. (c) 2D peak of graphene with a fit. (d) Conductivity ($\sigma$) versus back gate voltage ($V_G$) at 1.9 K along with a theoretical fit based on the self consistent theory for graphene.

**FIG. 2.** (a) Resistivity ($\rho$) versus back gate voltage ($V_G$) as a function of temperature ($T$). (b) $\rho$ versus $T$ at different $V_G$. (c) $\rho$ versus $V_G$ as a function of external magnetic field ($H$). (d) Magnetoresistance (MR) versus $V_G$ at 9 T.

**FIG. 3.** (a) Magnetoresistance (MR) versus external magnetic field ($H$) as a function of back gate voltage ($V_G$) along with two-fluid model fits. (b) The fit parameter $\alpha$ as function of $|V_G - V_D|$ derived both from experimental data and theory, which clearly shows the invalidity of this model. $V_D$ is a voltage corresponding to the CNP.

**FIG. 4.** (a) Magnetoresistance (MR) versus external magnetic field ($H$) at a back gate voltage ($V_G$) of 30 V which shows the signature of weak antilocalization at low magnetic fields (a sharp cusp). (b) The low field magnetoconductance versus $H$ as a function of $V_G$ with fits using the two-parameter empirical model. (c) Phase coherence length ($L_\phi$) as a function of $V_G$ at 1.9 K. (d) MR versus $H$ at different temperatures from another single layer graphene sample.

**FIG. 5.** (a) Resistance shows Shubnikov–de Haas oscillations(SdH) at 1.9 K and a normal magnetic field of 9 T. (b) Quantum Hall resistance with filling factors corresponding to Hall

plateaus is plotted as a function of back gate voltage ($V_G$). (c) Hall coefficient $R_H$ as a function of $V_G$ at 1.9 K. (d) Measured and estimated carrier concentrations ($n$) as a function of $V_G$ at 1.9 K.

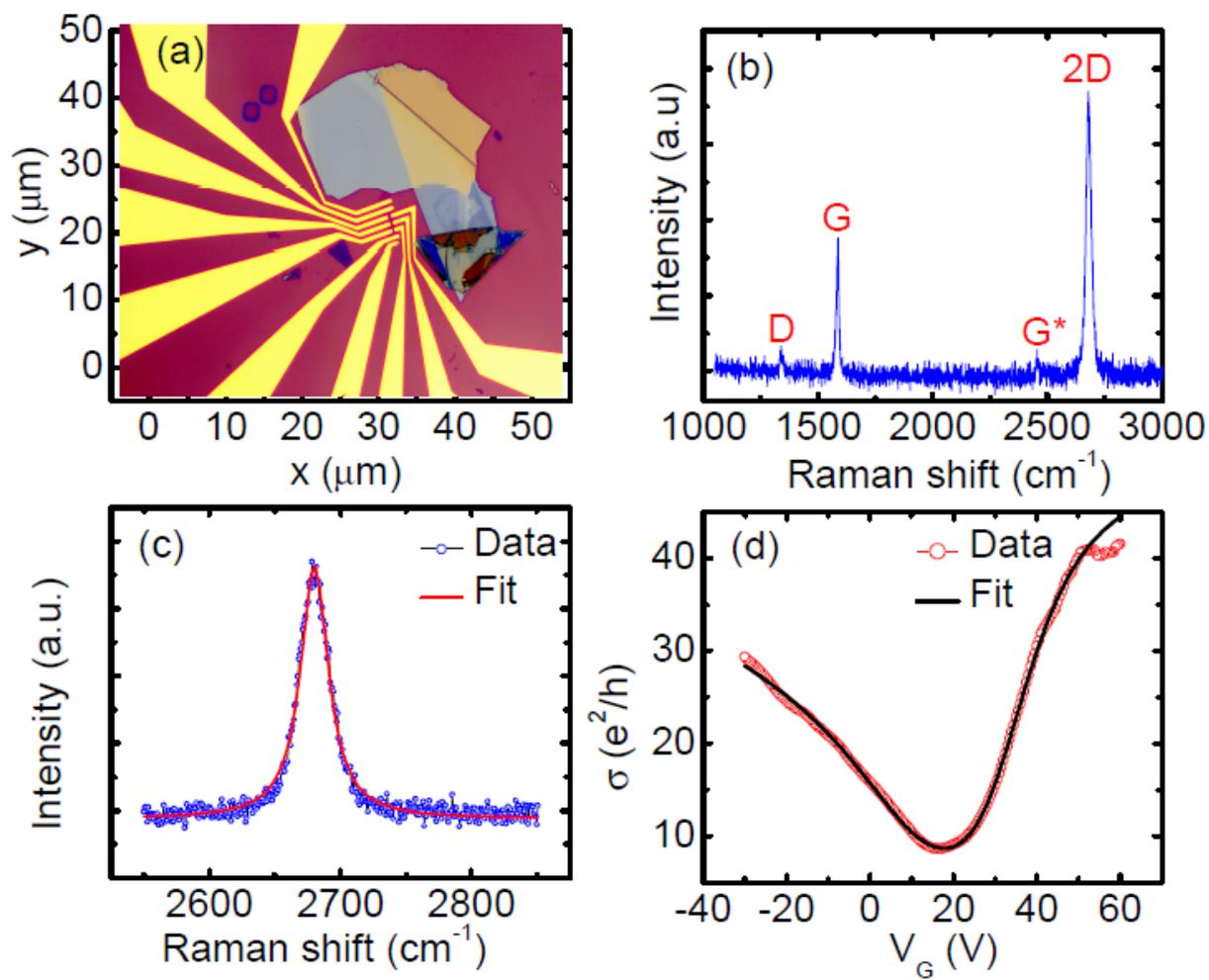

Figure 1
19

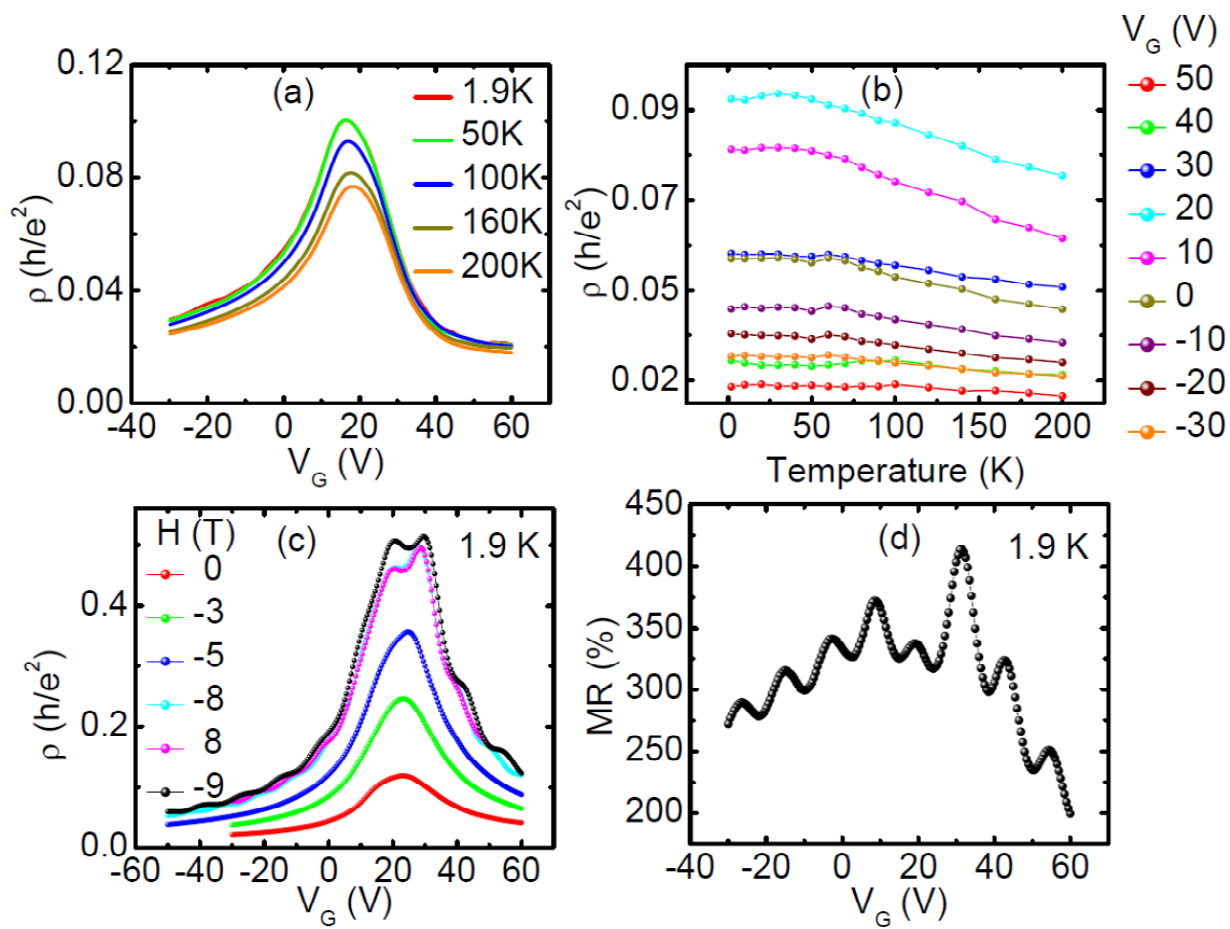

Figure 2

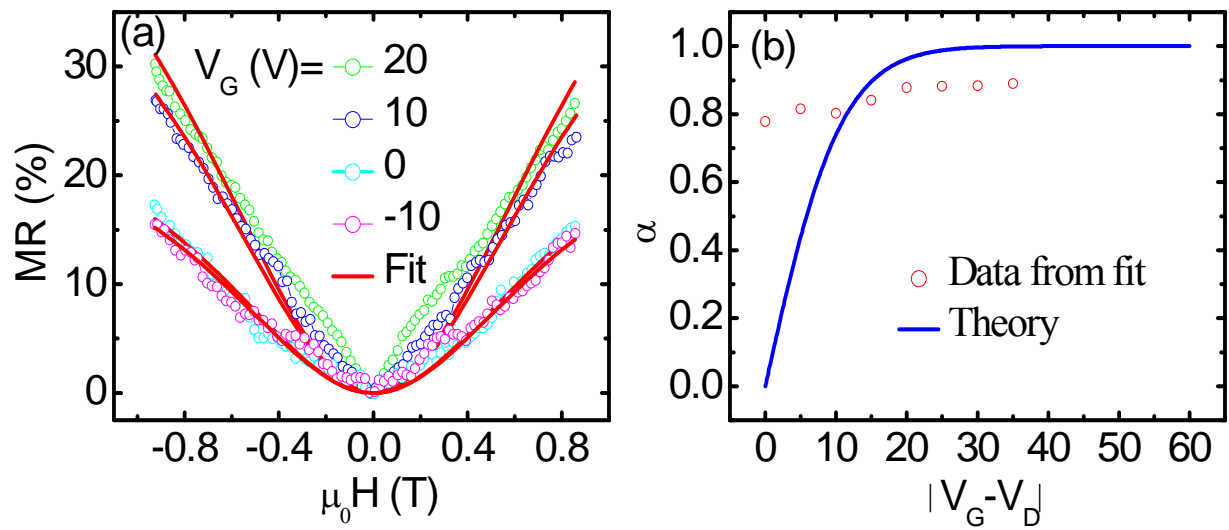

Figure 3

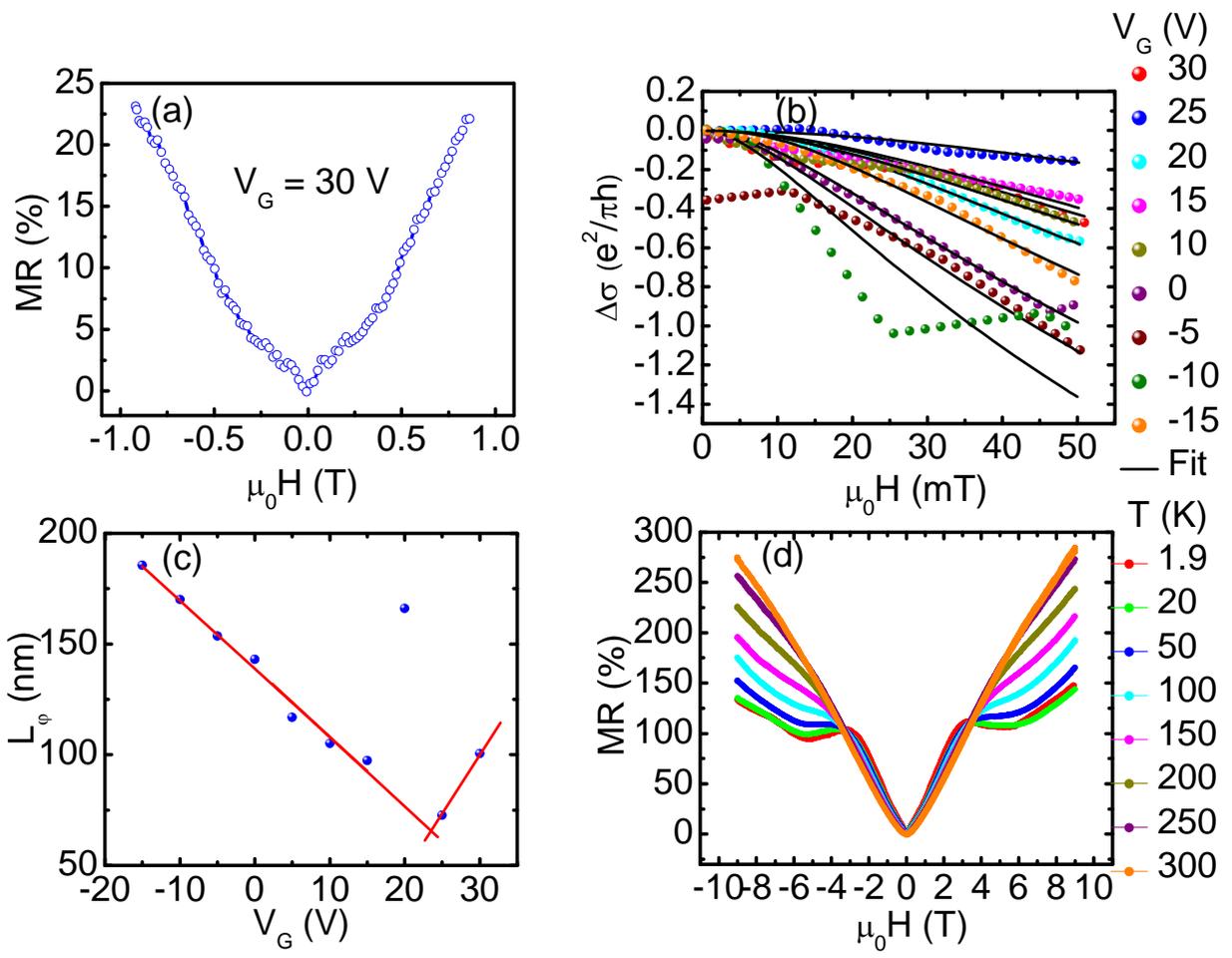

Figure 4

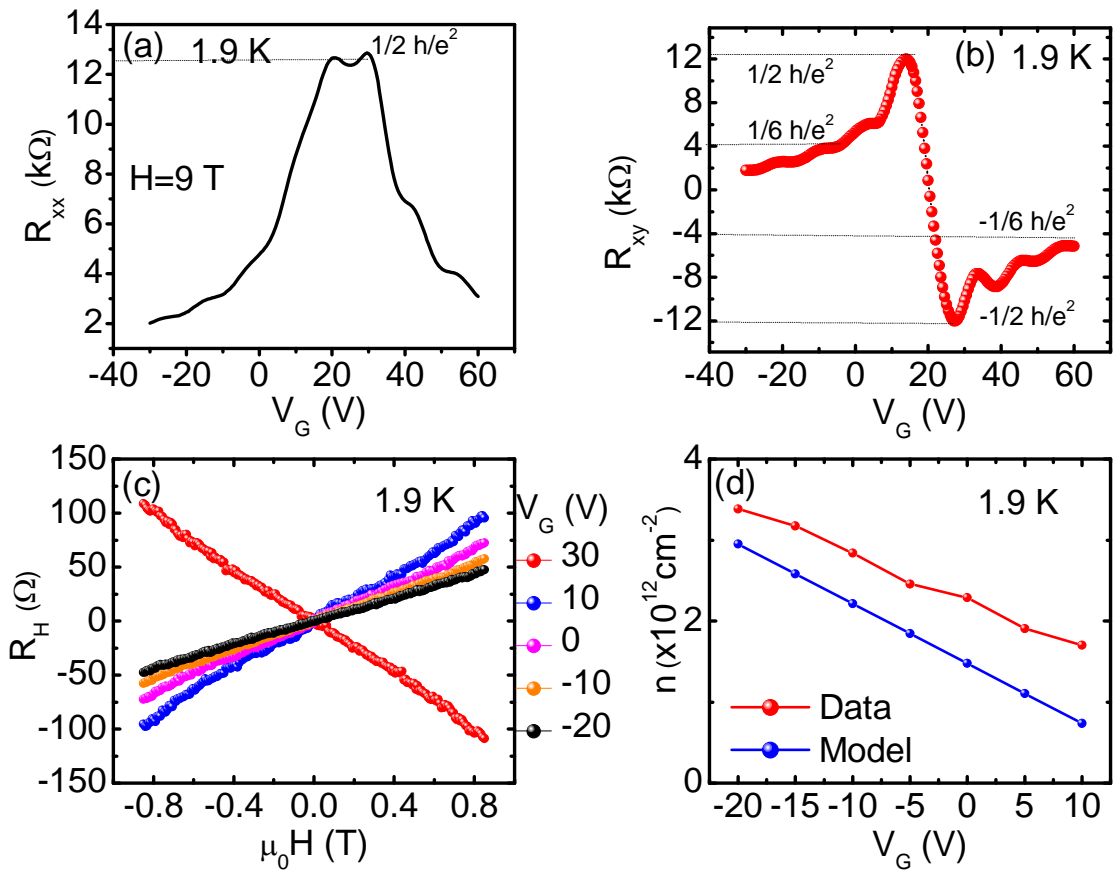

Figure 5

**Table 1. Parameters of the fittings using a phenomenological model.**

| $V_G(V)$ | $B_\varphi(T)$ | $B_*(T)$ | $L_\varphi(nm)$ | $L_*(nm)$ |
|---|---|---|---|---|
| 30 | 0.01629 | $1.7 \times 10^{-6}$ | 100.6 | 9935.1 |
| 25 | 0.03123 | $7.7 \times 10^{-6}$ | 72.7 | 4607.2 |
| 20 | 0.00597 | $1.8 \times 10^{-3}$ | 166.2 | 304.0 |
| 15 | 0.01735 | $1.0 \times 10^{-7}$ | 97.5 | 40161.1 |
| 10 | 0.01491 | $1.6 \times 10^{-5}$ | 105.1 | 3188.9 |
| 5 | 0.01207 | $4.2 \times 10^{-4}$ | 116.9 | 198.3 |
| 0 | 0.00805 | $1.3 \times 10^{-6}$ | 143.1 | 11066.5 |
| -5 | 0.00699 | $1.1 \times 10^{-7}$ | 153.6 | 39471.9 |
| -10 | 0.00569 | $5.1 \times 10^{-8}$ | 170.1 | 56835.3 |
| -15 | 0.00479 | $1.5 \times 10^{-4}$ | 185.6 | 331.9 |